\documentclass[10pt,twocolumn,english]{scrartcl}
\usepackage{lmodern}
\usepackage[T1]{fontenc}
\usepackage[latin9]{inputenc}
\usepackage{geometry}
\usepackage{units}
\usepackage{textcomp}
\usepackage{amsmath}
\usepackage{graphicx}

\makeatletter
\newcommand{\lyxaddress}[1]{
\par {\raggedright #1
\vspace{1.4em}
\noindent\par}
}

\@ifundefined{date}{}{\date{}}
\usepackage[font=sf,small, hang]{caption}
\makeatother

\usepackage{babel}
\begin{document}

\title{Functional imaging of ganglion and receptor cells in living human
retina by osmotic contrast}

\author{{\footnotesize{}Clara Pfäffle,}\textsuperscript{{\footnotesize{}1,2,{*}}}{\footnotesize{}
Dierck Hillmann,}\textsuperscript{{\footnotesize{}1,2,3,{*}}}{\footnotesize{}
Hendrik Spahr,}\textsuperscript{{\footnotesize{}1,2}}{\footnotesize{}
Lisa Kutzner,}\textsuperscript{{\footnotesize{}1,2}}{\footnotesize{}
Sazan Burhan,}\textsuperscript{{\footnotesize{}1}}{\footnotesize{}
Felix Hilge}\textsuperscript{{\footnotesize{}1}}{\footnotesize{}
}\\
{\footnotesize{}and Yoko Miura,}\textsuperscript{{\footnotesize{}1,2,4}}{\footnotesize{}
Gereon Hüttmann}\textsuperscript{{\footnotesize{}1,2,5}}}
\maketitle

\lyxaddress{{\scriptsize{}}\textsuperscript{1}{\scriptsize{} Institute of Biomedical
Optics, University of Lübeck, Peter-Monik-Weg 4, 23562 Lübeck, Germany}\\
{\scriptsize{}}\textsuperscript{2}{\scriptsize{} Medical Laser Center
Lübeck GmbH, Peter-Monik-Weg 4, 23562 Lübeck, Germany}\\
{\scriptsize{}}\textsuperscript{3 }{\scriptsize{}Thorlabs GmbH,
Maria-Goeppert-Straße 9, 23562 Lübeck, Germany}\\
{\scriptsize{}}\textsuperscript{4}{\scriptsize{}Department of Ophthalmology,
University of Lübeck, Ratzeburger Allee 160, 23562 Lübeck, Germany}\\
{\scriptsize{}}\textsuperscript{5}{\scriptsize{}Airway Research
Center North (ARCN), Member of the German Center of Lung Research
(DZL), 35392 Gießen, Germany}\\
{*}{\scriptsize{}Both authors contributed equally}}
\begin{abstract}
Imaging neuronal activity non-invasively in vivo is of tremendous
interest, but current imaging techniques lack either functional contrast
or necessary microscopic resolution. The retina is the only part of
the central nervous system (CNS) that allows us direct optical access.
Not only ophthalmic diseases, but also many degenerative disorders
of the CNS go along with pathological changes in the retina. Consequently,
functional analysis of retinal neurons could lead to an earlier and
better diagnosis and understanding of those diseases. Recently, we
showed that an activation of photoreceptor cells could be visualized
in humans using a phase sensitive evaluation of optical coherence
tomography data. The optical path length of the outer segments changes
by a few hundred nanometers in response to optical stimulation. Here,
we show simultaneous imaging of the activation of photoreceptor and
ganglion cells. The signals from the ganglion cells are ten-fold smaller
than those from the photoreceptor cells and were only visible using
new algorithms for suppressing motion artifacts. This allowed us to
generate a wiring diagram showing functional connections between photoreceptors
and ganglion cells. We present a theoretical model that explains the
observed intrinsic optical signals by osmotic volume changes, induced
by ion influx or efflux. Since all neuronal activity is associated
with ion fluxes, imaging osmotic induced size changes with nanometer
precision should visualize activation in any neuron. 
\end{abstract}

\section*{Introduction}

Observing and investigating the activity and wiring of the central
nervous system (CNS) in living humans can aid in a better understanding
of neuronal function. Anatomically and developmentally the retina
is part of the CNS. Therefore its neuron circuity, its specialized
immune response, and its blood-retina barrier resemble the respective
parts of the CNS \cite{KAUR,Streilein}. Given these similarities,
we may learn much about the CNS and peripheral nerves such as the
spinal cord by imaging the retina \cite{london}. Due to the optical
properties of the eye the retina is directly accessible to optical
imaging, with higher resolution (micrometer range) and more contrast
options than magnetic resonance imaging (MRI), functional MRI (fMRI),
or computed tomography (CT). This should allow a better diagnosis
not only of neurodegenerative ophthalmic diseases like glaucoma or
age-related macular degeneration (AMD), but also of diseases of the
CNS. For example, the progress of neurodegenerative diseases like
multiple sclerosis \cite{FISHER2006324,green,monteiro}, Parkinson's
disease, \cite{Devos1107,doi:10.1093/brain/awp068} or Alzheimer's
disease \cite{Danesh-Meyer1852,BLANKS1989364,BLANKS1996377,BLANKS1996385,PARISI20011860}
correlates with morphological changes in specific retinal regions.
Being the last neurons in the retinal circuity, which transmit the
visual information to the brain, the ganglion cells are of special
importance. Interestingly, they show morphological changes in all
above mentioned neurodegenerative disorders. In many of these, ocular
manifestation even precedes symptoms in the brain, therefore eye investigation
can offer earlier diagnosis \cite{london}. It is reasonable to assume
that, prior to morphological changes, neuronal function of the retina
is corrupted or changed. But currently we are lacking methods to objectively
check neuron function on a near cellular level in vivo. 

The main reason is that activation potentials of the neurons yield
only small optical changes \cite{Berlind,Hill426,Iwasa,OH} and are
therefore hard to detect. However, it has also been observed that
the volume of excited neurons increases due to osmotic processes \cite{Holthoff2,dmitriev,Huang,Holthoff1},
which elongate or shorten cell axes by tens of nanometers. Such volume
changes are orders of magnitude greater than optical path length changes
induced by changes of the refractive index by varying ion concentrations.
However, their detection is still challenging, because spatial changes
are far below the resolution limit of current clinically used imaging
methods like optical coherence tomography (OCT), ultrasound, or MRI.
Moreover optical path length (OPL) changes of inevitable eye motion
are orders of magnitude larger, corrupting measurements of osmotic
changes.

Volumetric phase-sensitive imaging can observe length changes in the
sub-wavelength range, provided that all motion related changes of
the retina are compensated. We recently detected the activity of human
photoreceptor cells (PRCs) after a light stimulus by using phase-sensitive
parallel OCT imaging of the whole field of view with a fast tunable
light source (Full-field swept-source OCT, FF-SS-OCT) \cite{Hillmann2016}.
The activation of the PRCs manifests itself in an elongation of the
OPL of their outer segments (OS). Shortly after that, measurements
by Zhang et al.~showed a similar behavior in mice \cite{Zhang}.
Based on these measurements, they introduced the idea that the observed
OPL elongation may be caused by an osmotically driven volume increase,
which mainly manifests in the axial dimension. Although an osmotically
driven process fits well to the dynamics of our observations, this
model leaves some questions unanswered: First, it is not obvious how
the concentration of osmotically active molecules increases inside
the photoreceptor OS to an extent that causes enough osmotic water
influx. Second, it is unclear how the plasma membrane could resist
the necessary dilation for the observed $10\,\%$ increase in length.
Finally, an inflow of water into the OS contradicts previous measurements
showing that photoactivation increases the volume of the extracellular
space (ECS) \cite{dmitriev,Huang}.

So far, in human retinas functional responses have only been observed
for the OS of photoreceptor cells, whereas no volume changes were
seen in retinal neurons, such as the ganglion cells. However, concentration
changes of ions causing osmotic effects are expected in these cells
and their ECS as well, which should result in measurable volume changes.

Here, we show that ganglion cells indeed exhibit an expansion similar
to the photoreceptor OS, albeit an order of magnitude smaller. By
improving the correction of residual motion artifacts, we were able
to simultaneously detect photoreceptor and ganglion cell activity.
These data allowed us to map the wiring of photoreceptors to ganglion
cells at different positions in the living human retina. Based on
our data, we provide an alternative theoretical model to explain the
observed change in photoreceptor OS and neuronal cells. This model
considers the limited stretchability of biological lipid membranes,
relies on the molecular concentration change of ions during the hyperpolarisation,
and is consistent with reported volume changes of the ECS. 

\section*{Results and Discussion}

\begin{figure*}
\centering{}a)\includegraphics[width=0.3\textwidth]{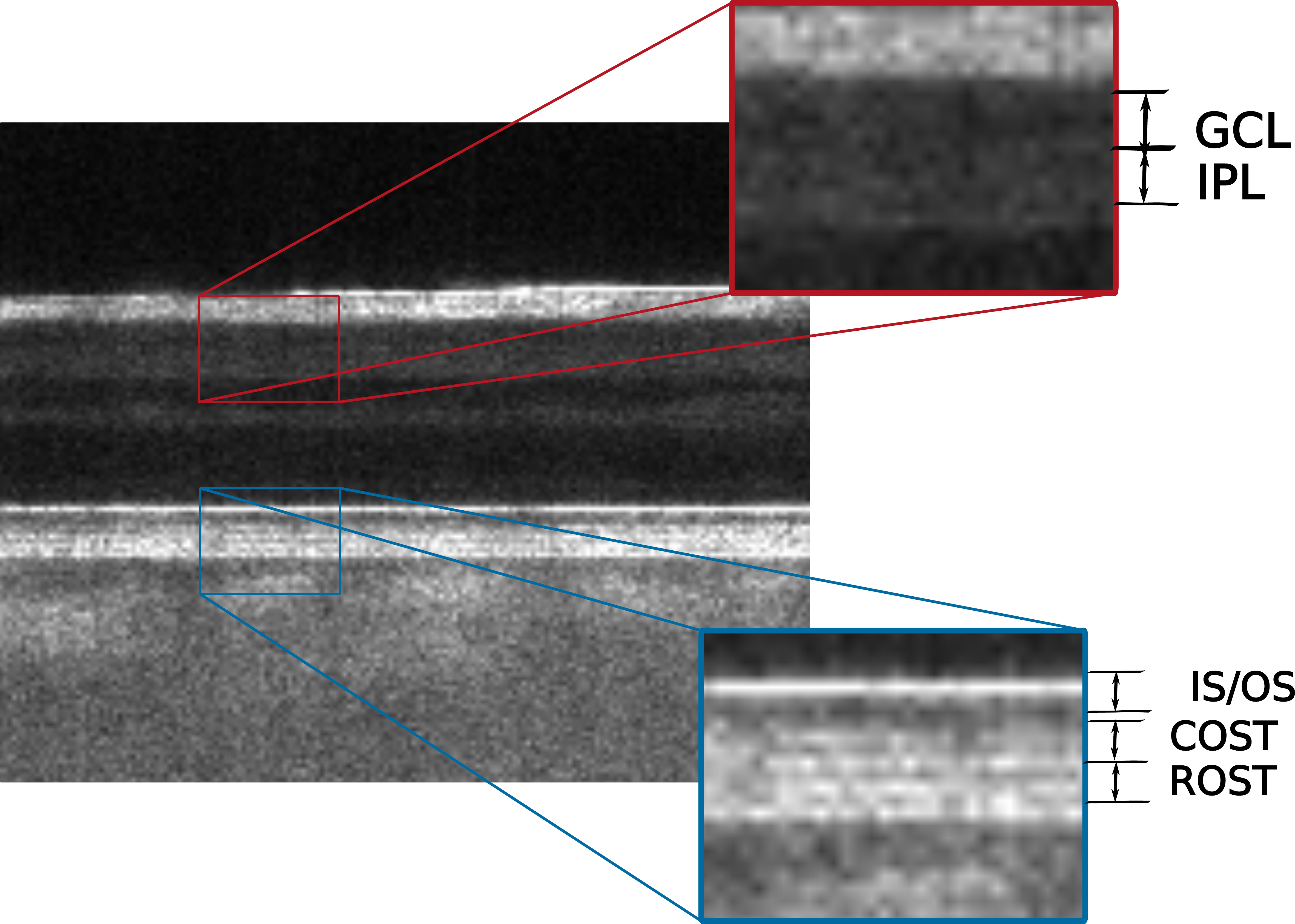}
b)\includegraphics{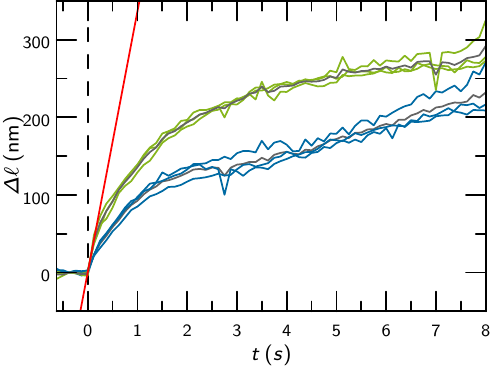}c)\includegraphics[width=0.3\textwidth]{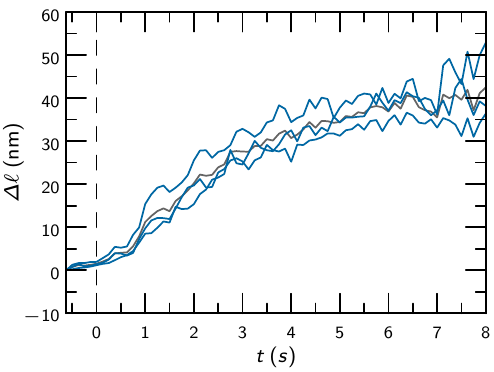}\caption{{\footnotesize{}a)~B-scan of the retina in $3\:\mathrm{mm}$ distance
from the center of the macula. For the OPL of the cone and the rod
OS, phase difference of the IS/OS junction (IS/OS, averaged over 2\ px)
and the cone OS tips (COST, averaged over $2\,\mathrm{px}$) or the
rod OS tips (ROST, averaged over $2\,\mathrm{px}$) were calculated,
respectively. For the OPL changes of the ganglion cells the phase
differences between the GCL (GCL, averaged over $7\,\mathrm{px}$)
and the IPL (blue, averaged over $5\,\mathrm{px}$) were determined.
b) Optical path length (OPL) changes of cone (blue) and rod (green)
OS. The averaged time course of those measurements are shown in grey.
Initially, the rod OS elongates with a rate of $336\,\mathrm{nm/\mathrm{s}}$,
as shown in red. c) OPL changes between GCL and IPL, which describes
the optical distance of the ganglion cell bodies to their synapses
for three different measurements (blue) and its averaged (black).
All OPL changes were measured during 8 s stimulation in 3 mm distance
from the center of the macula.}}
\label{time_course}
\end{figure*}

The achievable axial resolution of OCT images is limited by the bandwidth
of the light source and is typically a few micrometers. But in addition
to the image intensity, the phases of the backscattered light are
also acquired by interference and are sensitive to nanometer distance
changes. However, in most OCT systems the phase information are corrupted
by motion or scanning artifacts which render them worthless. Our FF-SS-OCT
guaranteed phase stability and therefore allowed us to detect changes
of the sample in the nanometer range. To image minute changes in the
length of cone and rod OS, the phase differences between the inner
segment outer segment junction (IS/OS) and the tips of the cone and
rod OS were calculated, respectively. To detect changes of the OPL
in the ganglion cells, we computed the phase difference between the
ganglion cell layer (GCL), which contains the ganglion cell bodies,
and the inner plexiform layer (IPL) with the synaptic connections
from the ganglion cells with the bipolar and amacrine cells (Fig\ \ref{time_course}a).
To measure smaller OPL changes on a background of more severe motion
artifacts, as present in the GCL, we needed robust and sub-pixel precise
post processing of the volumetric OCT data; the critical areas for
the processing were segmentation, co-registration, correction of pulsation
artifacts, and referencing the phase time course to a background signal. 

With a sub-pixel precise segmentation and co-registration (see Methods),
we were able to obtain phase stable measurements over more than $8\,\mathrm{s}$.
Afterwards, the first 5 volumes, in which no stimulation occurs and
which correspond to $625\,\mathrm{ms}$, were used for a dynamic phase
referencing with a volume that was recorded in the same phase of the
heart beat. This approach reduced motion artifacts due to the pulsation
\cite{Spahr2015}. After dynamic phase referencing, the phase differences
of the GCL and the IPL was still dominated by non periodic motion.
However, averaging the phase differences over the whole stimulation
time led to distinct recognizable stimulated areas. When extracting
the phase time courses of those areas, the remaining motion artifacts
were removed by averaging the phase difference of the background and
subtracting it from the phase difference in the activated area. This
post processing could finally extract time courses of the intrinsic
optical signals (IOSs) between the GCL and the IPL .

\begin{figure*}
\centering{}\includegraphics[width=0.9\textwidth]{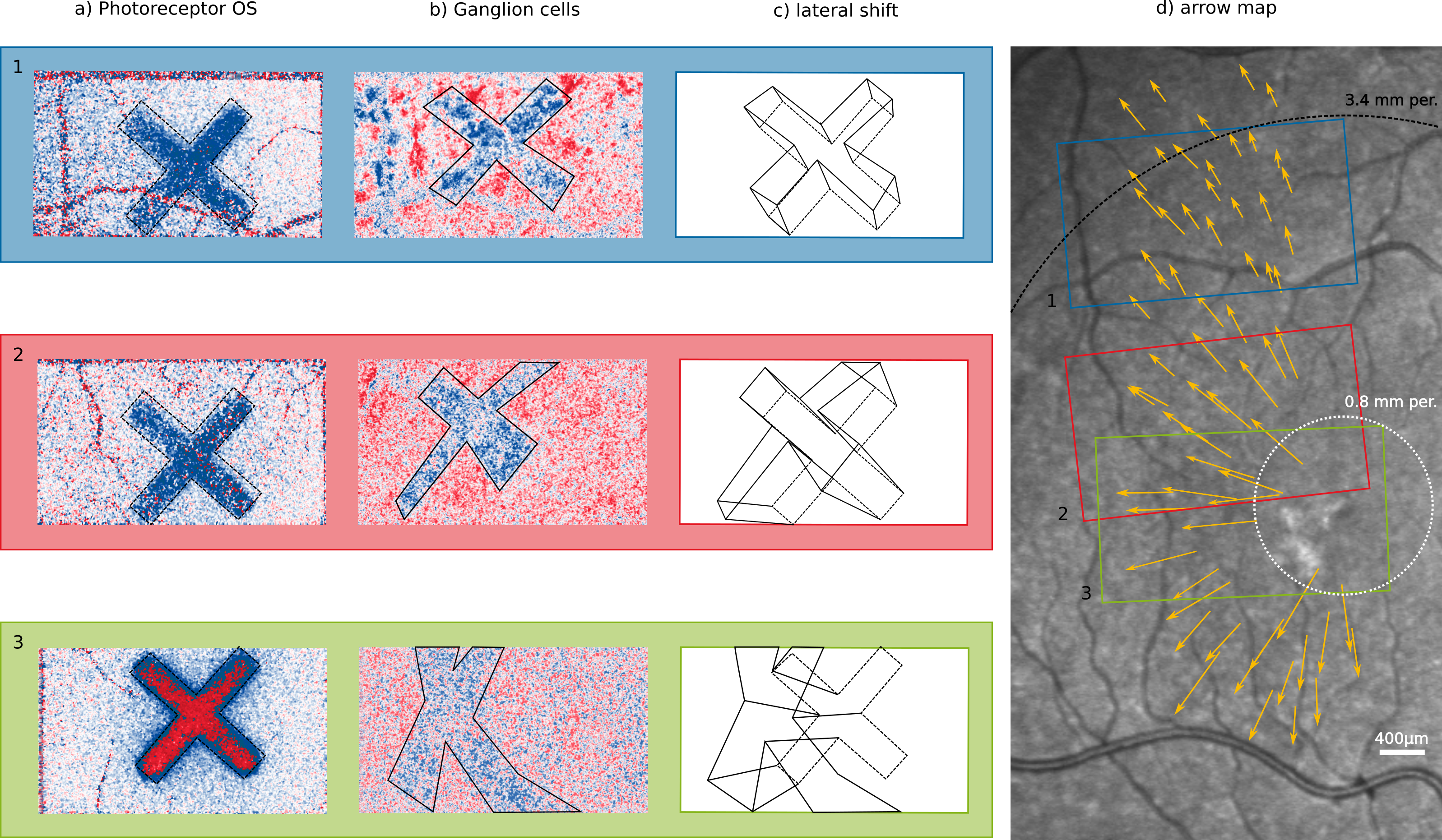}\caption{{\footnotesize{}IOS pattern for different positions in the retina.
Top row:~in $3.5\,\mathrm{mm}$ superior temporal, center row:~in
$2\,\mathrm{mm}$ superior temporal and bottom row:~in $1.8\,\mathrm{mm}$
temporal from the fovea. The positions correspond to the marked positions
in the fundus image at the right. Column a)~shows the observed pattern
in the photoreceptor OS for each position and column b) in the GCL.
Each pattern was outlined manually (Column c) and the corresponding
edge points in the different layers were linked to generate the arrow
map (Column d). An SLO image (Heidelberg engineering) of the subject
is laid underneath the arrow map. The white and black circles mark
the positions in $0.8\text{\,\textmu m}$ and $3.4\,\text{\textmu m}$
from the center of the macula, respectively.}}
\label{translation}
\end{figure*}

As observed for the photoreceptor OS, the GCL shows an increase of
the optical path length in the stimulated area. It is, however, an
order of magnitude smaller. The increase in OPL of the GCL reached
its maximum of about $40\,\mathrm{nm}$ after approximately $5\,\mathrm{s}$
(Fig.~\ref{time_course}a), whereas the elongation of the OS did
not reach a saturation state after $8\,\mathrm{s}$ stimulation at
a length dilatation of more than $290\,\text{\ensuremath{\mathrm{nm}}}$
for rods and $240\,\mathrm{nm}$ for cones (Fig.\ \ref{time_course}b).
As expected, the activated area of the ganglion cells is laterally
shifted with respect to the activated area of the photoreceptor OS
as shown in Fig.~\ref{translation}. In addition to the lateral shift,
a deformation of the activated area occurred. Both lateral shift and
deformation are highest in direct neighborhood of the fovea. The mapped
shift of the IOS shows a characteristic translation pattern (Fig.\ \ref{translation}),
the direction of this shift points radially outwards from the fovea,
as was histologically shown before ex vivo by Drasdo et al. \cite{DRASDO}.
A larger shift near the fovea attributes to the absence of ganglion
cells and the high density of cones in the fovea. In the case of our
subject, the largest displacement is about $650\,\mathrm{\text{\textmu}m}$
at $0.8\,\mathrm{mm}$ temporal, which agrees well with the results
of Drasdo et al. ($406-632\,\text{\textmu m}$ at $0.85-1.348\,\mathrm{mm}$
temporal); in $3\,\mathrm{mm}$ from the center of the fovea the shift
is only about $180\,\mathrm{\text{\textmu m.}}$ For the same stimulated
area the activated area in the GCL is therefore larger near the fovea,
because the receptive fields are smaller and each cone is connected
to one individual ganglion cell, while in the periphery cone density
decreases and the ganglion cells process information from several
rods.

As it was suggested before \cite{Zhang} an osmotically driven volume
change is a fitting explanation for the OPL elongation over hundreds
of nanometers within several seconds. However, the largest concentration
changes in photoreceptor cells and neurons are associated to changes
of the membrane potential. During excitation, the neuron alters the
permeability of its plasma membrane for specific ions, in order to
shift the electrochemical equilibrium by a net influx of cations.
This polarizes the membrane potential from negative to positive. But
this rapid shift also changes the osmolarity in the cell and in the
ECS, and therefore should lead to an osmotic water influx into the
neuron. In contrast to neurons, stimulation of photoreceptor cells
does not result in a depolarisation but in a hyperpolarisation of
the plasma membrane \cite{yau}. The hyperpolarisation is caused by
an inhibition of a dark current, which is caused by a permanent influx
of cations into the cell. This dark current is compensated by an active
transport of ions over the plasma membrane of the inner segment back
into the ECS. The inhibition of the dark current therefore leads to
a net efflux of cations into the ECS, which decreases the osmolarity
of the photoreceptor cells and therefore should lead to a water efflux
from the cells into the ECS. This different osmotic behavior of photoreceptor
and ganglion cells after a light stimulus was already shown by Dmitriev
et al.~in chick retina \cite{dmitriev}. They measured a volume increase
of the ECS around the photoreceptor cells, reaching its maximum after
approximately 30~s; at the same time they observed the volume of
the ECS decreasing around bipolar, amacrine, and ganglion cells, reaching
saturation after about 5~s. In the ganglion cells, the measured IOS
approach a saturation within 8\ s of measurement time, too. Unfortunately,
we could not generate reach phase stability over a longer measurement
time and therefore did not observe a saturation of the photoreceptor
elongation. Nevertheless, the general time courses of the photoreceptors
and especially of the GCL that we obtained fit well to the ECS volume
changes that Dmitriev et al. measured in the respective layers of
chick retina.

At first glance, the postulated decrease of the OS volume contradicts
an elongation of the photoreceptor OS. Indeed, the length of isolated
OS shrinks by up to 40\% under unphysiologically strong hypertonic
conditions \cite{Korenbrot20}, but these measurements can arguably
be transferred to physiological scenarios. The opposite, a significant
elongation of the OS by an expanding volume would only be possible
if accompanied by a significant increase of the membrane surface,
but biological membranes are inelastic and rupture at about 3\% dilation
of their surface \cite{sackmann}. Furthermore, Cohen et.~al observed
that strong hypotonic conditions lead to a bulging of the OS, by which
the diameter is increased and the length decreased \cite{Cohen}.
Therefore, it is unlikely that cells compensate volume increase by
a dilation of their plasma membrane. Instead, it is more likely that
any physiological volume change is compensated by conformation, as,
e.g., known from vesicles \cite{Dobereiner}. Following, we will show
that the assumption of an osmotically driven volume decrease at constant
and smooth surface area explains qualitatively and quantitatively
the observed elongation of rod OS. 

\begin{figure}
\centering{}\includegraphics[width=0.95\columnwidth]{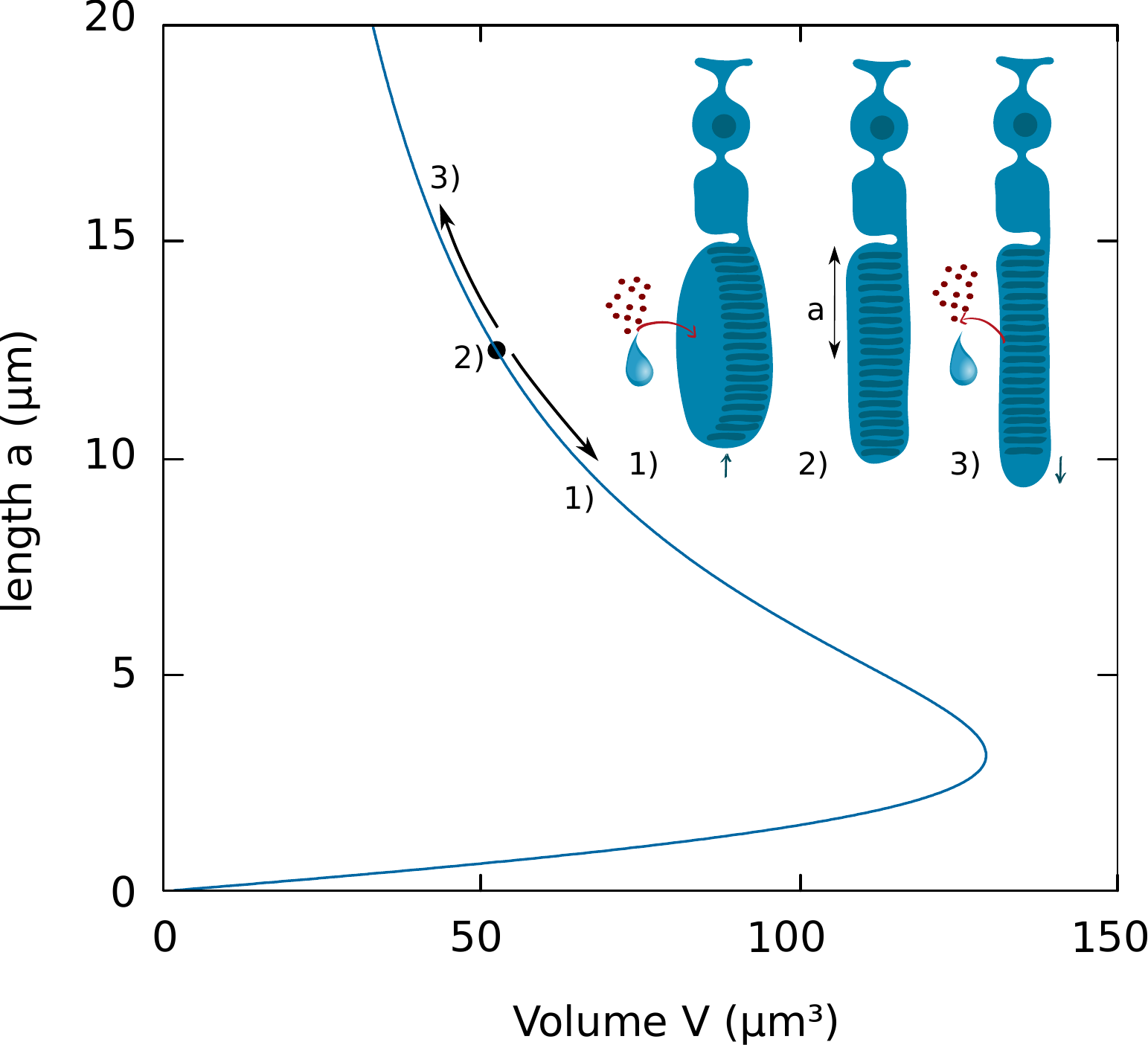}\caption{{\footnotesize{}Volume $V$ of biaxial ellipsoids with constant surface
area ($124\,\text{\ensuremath{\mathrm{\text{\textmu m\texttwosuperior}}}}$)
as the length of its rotational axis $a$ varies. Position (2) resembles
the starting conditions of PRC OS with a vertical radius $a=12.5\,\mathrm{\text{\textmu}m}$
and volume of $52.4\,\mathrm{\text{\textmu}m^{3}}$. Position (1):
~Decreasing the radius $a$ leads to an increased volume. Position
(3): Similarly, increasing the radius $a$ decreases leads to a decreased
volume. }}
\label{volume_vs_axis}
\end{figure}

For simplicity we consider only the rod OS and assume a constant surface
area. Since photoreceptor discs show no effect under different osmotic
conditions \cite{Cohen}, we expect cones to behave similarly, although
we expect the kinetics of the process to be more complex. We describe
the conically shaped rod OS as a biaxial ellipsoid, where the initial
conformation has one long radius $a=12.5\,\text{\textmu m}$ and two
short radii $b=1\,\text{\textmu m}$ \cite{kraft}. The area surface
of the ellipsoid is given by 
\[
A_{\mathrm{ROS}}=2\pi b\left(b+\frac{a^{2}}{\sqrt{a^{2}-b^{2}}}\arcsin\left(\frac{\sqrt{a^{2}-b^{2}}}{a}\right)\right)
\]
and its volume is computed to
\[
V_{\mathrm{ROS}}=\frac{4}{3}\pi ab^{2}\text{.}
\]
The surface area of the rod OS, $A_{\mathrm{ROS}}$, computes to $124\,\mathrm{\text{\textmu m}}{{}^2}$;
its initial volume to $V_{\mathrm{ROS}}=52.4\,\mathrm{\mathrm{\text{\textmu m\textthreesuperior}}}$.
Assuming the surface area $A_{\mathrm{ROS}}$ to be constant, elongating
the long axis $a$ will shrink the short axis $b$ and thereby decrease
the volume. In contrast, increasing the volume of the outer segment
would shrink the long and elongate the short axis, thereby increasing
the volume-to-surface ratio by getting closer to a spherical shape
\cite{Cohen}(Fig.$\:$\ref{volume_vs_axis}). 

To quantitatively characterize the process we consider the initial
slope of the elongation of the OS, since at this point no saturation
or other effects interfere with the dynamics of the process. The initial
slope of the rod OS elongation $\ell$ is $\mathrm{\frac{\mathrm{d}\ell}{\mathrm{d}t}}=336\,\mathrm{\mathrm{\frac{\mathrm{nm}}{\mathrm{s}}}}$;
given a refractive index $n$ of $1.41$ \cite{Sidman15}, the long
radius of the OS increases by $\frac{\mathrm{d}a}{\mathrm{d}t}=\frac{1}{2n}\frac{\mathrm{d\ell}}{\mathrm{d}t}=119\,\mathrm{\text{\ensuremath{\frac{\mathrm{nm}}{\mathrm{s}}}}}$.
Assuming a constant surface area this results in a decreases of the
short axis by $\frac{\mathrm{d}b}{\mathrm{d}t}=9.5\,\mathrm{\frac{\mathrm{nm}}{\mathrm{s}}}$
and a volume decrease with a rate of $\frac{\mathrm{d}V_{\mathrm{ROS}}}{\mathrm{d}t}=0.5\,\mathrm{\text{\ensuremath{\frac{\mathrm{\text{\textmu m\textthreesuperior}}}{\mathrm{s}}}}}$. 

To check if this volume change is consistent with the concentration
change during the excitation of the PRCs, we need to determine the
ion flux due to the stimulation. In our experiments, the stimulus
intensity $P$ is $27.5\,\mathrm{\text{\textmu}W}$ at a central wavelength
$\lambda_{0}$ of $550\,\mathrm{nm}$ and illuminates an area $A_{\text{stimulus}}$
of about $1.3\:\mathrm{mm\text{\texttwosuperior}}$ on the retina.
For a mean photon energy $E=\frac{h\mathrm{c}}{\mathrm{\lambda_{0}}}$
, with $h$ being the Planck constant and $\mathrm{c}$ being the
speed of light, this corresponds to a photocurrent density of $\nicefrac{P}{EA}=5.86\times10^{7}\text{s\textsuperscript{-}\textonesuperior\,\textmu m\textsuperscript{-}\texttwosuperior}$.
After correcting for corneal reflection (4\%), ocular media absorption
(50\%), and photons passed through the retina without being absorbed
(80\%) \cite{hecht}, $5.6\times10^{6}\,\text{s\textsuperscript{-}\textonesuperior\textmu m\textsuperscript{-}\texttwosuperior}$
photons remain for activation of the photoreceptor OS. This photon
flux saturates the rods and therefore yields the maximum ion current,
which is $I_{\mathrm{max}}=13\,\mathrm{pA}$ in human rods \cite{kraft}.
It is caused by the stimulus-induced inhibition of the dark-current,
which is mainly composed of single charged Na\textsuperscript{+} ($r_{\mathrm{Na^{+}}}=80\%$),
but also double charged Ca\textsuperscript{2+} ($r_{\mathrm{Ca}^{2+}}=15\%$)
and Mg\textsuperscript{2+} ($r_{\mathrm{Mg}^{2+}}=5\%$) \cite{yau}.
The resulting concentration change rate $\mathrm{\frac{\mathrm{d}c_{ion}}{\mathrm{d}t}}$
in the photoreceptor OS is thus computed to

\[
\frac{\mathrm{d}c_{\mathrm{ion}}}{\mathrm{d}t}=\frac{I_{\mathrm{max}}}{e\cdot N_{\mathrm{A}}\cdot0.5\cdot V_{\mathrm{\mathrm{ROS}}}\cdot\left(r_{\mathrm{Na^{+}}}+2r_{\mathrm{Ca}^{2+}}+2r_{\mathrm{Mg}^{2+}}\right)}\text{,}
\]
with $e$ being the elementary charge, $N_{\mathrm{A}}$ the Avogadro
constant, and $V_{\mathrm{ROS}}$ the volume of the rod OS. Due to
the dense packaging with discs only half of the volume is filled with
solvent, accounting for the factor 0.5 in the formula. The ion concentration
increases by $4.28\,\mathrm{mOsM/s}$, corresponding to $1.42\,\frac{\%}{\mathrm{s}}$
of the physiological osmolarity of $300\,\mathrm{mOsM}$ \cite{BardyE2725}.
According to Van't Hoff law and assuming this concentration change
is completely compensated by a volume change, this results in a rate
of volume change rate of the same relative size ($\frac{\mathrm{d}V}{\mathrm{d}t}=1.42\,\frac{\%}{\mathrm{s}}\times0.5\cdot V_{\mathrm{ROS}}=0.374\,\mathrm{\frac{\mathrm{\text{\textmu m\textthreesuperior}}}{\mathrm{s}}}$),
which is close to the value we calculated based on the observed length
change in the OS ($0.5\,\mathrm{\frac{\mathrm{\text{\textmu m\textthreesuperior}}}{s}}$). 

For the ganglion cells we expect an opposite volume change, since
a decrease of the ECS volume was observed by Dmitriev et al. The phase
evaluation, however, shows an elongation of the OPL in both layers.
Here the situation is less clear than for the OS for two reasons:
First, since the ganglion cells have a nearly spherical cell body,
it is difficult to predict in which direction the cells will elongate
in case of a volume increase. And second, the path length changes
were measured not only over the cell bodies but included the synaptic
endings of the ganglion cells. Therefore the phase changes do not
necessarily correspond to an elongation of the cell bodies, but may
also be explained by an expanding distance of these two structures.

Still, the general time course, magnitude, and saturation of the IOS
in the GCL fit well to the observed volume change of Dmitriev et al.
\cite{dmitriev}. For this reason, we are confident that the OPL changes
are also osmotically driven conformation changes of the ganglion cells.

\section*{Conclusion}

Here, we demonstrate the non-invasive, simultaneous measurements of
photoreceptor and ganglion cell activity after optical stimulation
with high spatial and temporal resolution in the living human eye.
Currently, we lack sensitivity as well as lateral and axial resolution
to resolve single ganglion cell activity. However, as technology for
phase stable imaging on a cellular level improves \cite{Liu12803}
we expect investigations of single ganglion cell behavior in the living
human eye to become possible in the near future. This will allow studying
of neuron networks and information processing in the retina and even
the brain. Future investigations will also have to show the value
of this functional imaging for clinical diagnostics. Given the increasing
clinical use of subtle morphological changes in the retina for diagnosing
neuronal disorders, it is well possible that functional single neuron
imaging will turn to an important tool there.

Our theoretical model gives an explanation for the molecular origin
of our signals, which is consistent with earlier investigation of
the PRC regarding concentration and volume changes. Nevertheless,
we are aware that this explanation is currently based on a simple
theoretical model and needs further experimental validation. The analysis
of our results suggests that neuronal activity is visible by osmotic
fluid changes between the ECS and the neuronal cell as it was observed
earlier in chick eyes by Dmitriev et al \cite{dmitriev}. Consequently,
using osmotic contrast we should be able to observe activity in the
bipolar and amacrine cell layers as well. One of our next tasks will
be the visualization of OPL changes in this region to complete the
functional imaging of the neuronal retina. 

\subsubsection*{Acknowledgments}

This research was funded by the German Research Foundation (DFG),
project Holo-OCT HU 629/6-1.

\subsubsection*{Competing interests}

D.H. works at Thorlabs GmbH, which manufactures and sells OCT devices. 

\section*{Material and Methods}

\paragraph{Setup and data acquisition}

\begin{figure}
\centering{}\includegraphics[width=0.9\columnwidth]{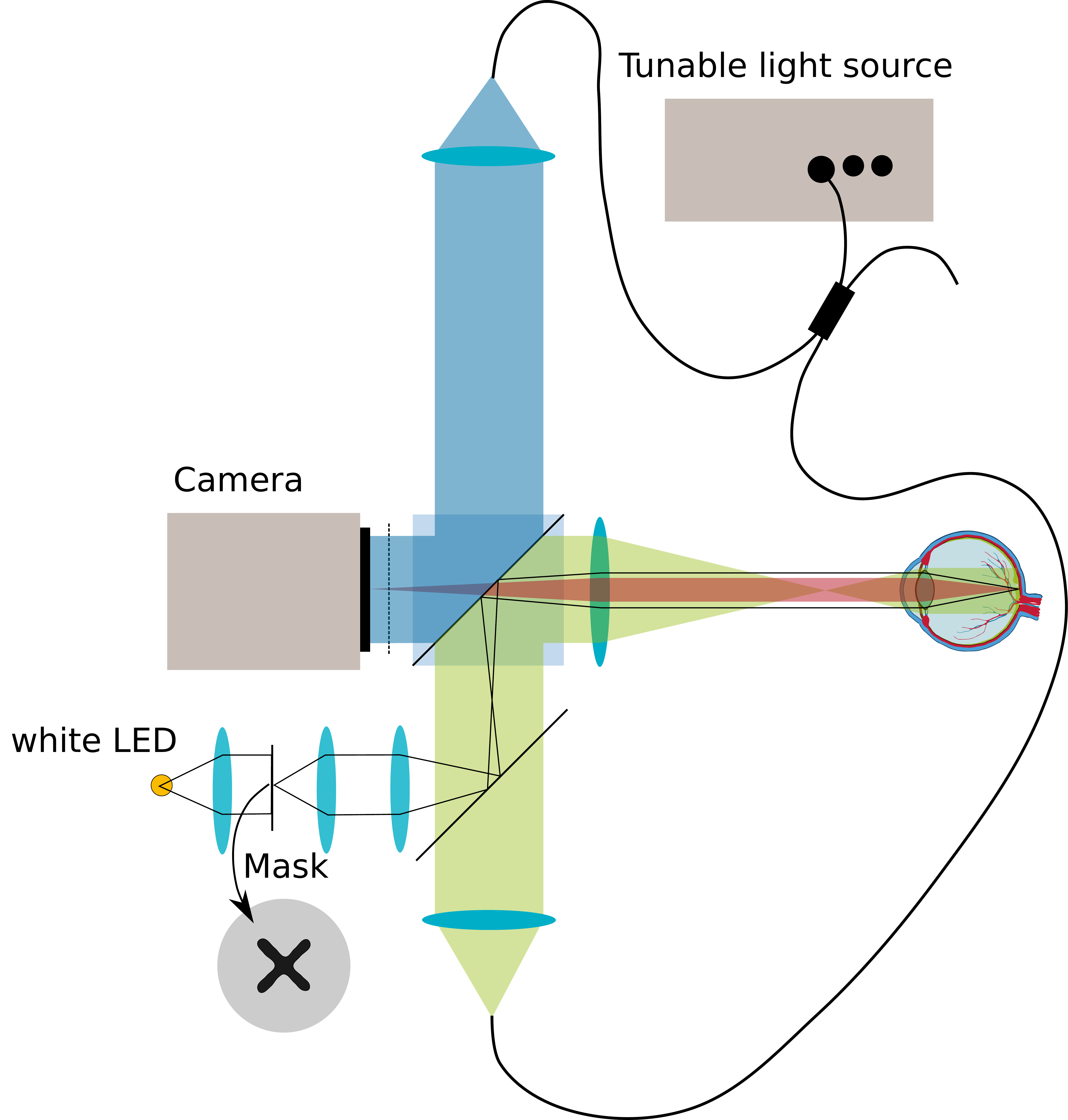}\caption{{\footnotesize{}Setup for imaging photoreceptor and ganglion cell
activity. The retina of the subject is illuminated with a collimated
beam (green) and then imaged onto the sensor of a high-speed 2D camera,
where it is superimposed with a collimated reference beam (blue).
The stimulation of the retina is done by a white LED with an 'x'-shaped
mask in the image plane, coupled into the sample illumination via
a cold light mirror.}}
\label{Set-up}
\end{figure}

The retina was imaged with a full-field swept-source OCT (FF-SS-OCT)
system based on a Mach-Zehnder type interferometer (Fig.~\ref{Set-up}).
The light of a swept source (Superlum BroadSweeper BS 840-1, central
wavelength 841.5~nm, 51~nm sweep range) was split into reference
and sample beam. The reference beam was collimated and brought onto
the sensor of a high-speed camera (FASTCAM SA-Z, Photron). The sample
beam illuminated the retina with a parallel beam at an irradiation
power of $5.2\,\mathrm{mW}$. The backscattered light was imaged onto
the camera, where it was superimposed with the reference beam. The
central $640\times368$ pixels of the camera were read out at a frame
rate of $60\,\mathrm{kHz}$. During one wavelength sweep, 512 images
were recorded to acquire one volume in $8.5\,\mathrm{ms}$. This corresponds
to an A-scan rate of $27.6\,\mathrm{MHz}$. With these parameters,
70 volumes could be acquired until the memory of the camera was exhausted,
corresponding to a total measuring time of $0.595\,\mathrm{s}$ at
full duty cycle. For longer measurements, the acquisition of a volume
was only triggered every $125\,\mathrm{ms}$, which enabled a total
measurement time of $8.75\,\mathrm{s}$. Although longer measurements
are theoretically possible with this setup, we could not maintain
phase stability for longer times, because speckle patterns and phases
decorrelated too strongly.

The volumes were first reconstructed from the camera data as described
in previous publications \cite{Hillmann2016}. While lateral phase
stability in one volume was achieved by the parallel imaging of all
lateral positions, the axial phase error was effectively corrected
via an optimization of image quality \cite{HillmannAberr}. To achieve
phase stability of the OCT data between volumes, their positions were
aligned with sub-pixel precision to cancel phase changes due to bulk
motion. This was achieved by a suitable co-registration and segmentation
of the acquired volumes. For the stimulation of the retina white light
was used, which was coupled into the sample beam via a cold mirror.
A mask illuminated by the white LED was imaged onto the retina leading
to an 'x'-shaped stimulation pattern, Total irradiation power was
$27.5\,\mathrm{\text{\textmu W}.}$ For more detailed description
of the setup see \cite{Hillmann2016}

All investigations were done with fully dark adapted healthy volunteers
with a medically dilated pupil of $8\,\mathrm{mm}$; written informed
consent was obtained from all subjects. Compliance with the maximum
permissible exposure (MPE) of the retina and all relevant safety rules
was confirmed by the safety officer. The study was approved by the
ethics board of the University of Lübeck (ethics approval Ethik-Kommission
Lübeck 16-080)

\paragraph{Co-registration}

To co-register the volumes, three major steps were performed: A lateral
coarse registration, a lateral fine registration, and an axial registration.
For each step, we computed the deformation with respect to a master
volume and individually corrected each volume.

For the lateral coarse registration, we laterally divided the volumes
into tiles of size $256\times256$ pixels. The magnitude of the complex-valued
OCT signal of each tile were then Fourier transformed in $z$-direction
and again their magnitude was computed. This step removed any dependence
of the $z$-position, which would otherwise be encountered as a $z$-dependent
phase due to the Fourier shift theorem. Additionally, we cut off the
lower $10\%$ of all axial frequencies to remove high-frequency noise.
These data were then phase correlated to determine the lateral $x$
and $y$ displacement between the respective tiles of two volumes
that are compared. Finally, the median of all resulting $x$ and $y$
displacements between the respective and the master volume were taken
to determine the coarse displacement. 

The lateral fine registration was performed to also take deformations
of the volumes into account. For this, tiles of size $128\,\times128\,$pixels
were created. Their correlation was computed in analogy to the previous
step, but instead of taking the median, the displacements were interpolated
to the full volume to give a displacement map as a function of the
lateral position. To correct the displacements, an algorithm for a
fast Fourier transform on non-equispaced data \cite{potts} in $x$
and $y$ direction was used to compute the correctly interpolated
Fourier representation for all depth layers of the volume. A standard
FFT to get back to position space gave interpolated results. 

Finally, to obtain an axial displacement map, we divided into tiles
of $96\times96$ pixels. The point-by-point magnitudes of the complex-valued
tiles were phase correlated and the maximum of the phase correlation
gave a $z$-displacement between the respective tiles of the two volumes.
An interpolation upscaled the $z$-displacement map to give suitable
values for all lateral points. Finally, we used these values to shift
each A-scan in each volume axially by applying the Fourier shift theorem. 

\paragraph{Segmentation}

Segmentation of FF-SS-OCT data with sub-pixel precision, faced two
major challenges. First of all, the FF-SS-OCT images suffer from poor
signal-to-noise ratio (SNR), and second, the data size is huge ($70$
volumes, $640\,\times368\,\times256\,$pixels each), leading to long
computation times. Therefore an algorithm was needed that is robust
against noise and evaluates in reasonable time. We achieved this by
a combination of coarse graining, diffusion maps \cite{coifman},
and $k$-means clustering; the basic idea for this was previously
proposed by Raheleh Kafieh et al.~\cite{Dmaps}. After co-registering
all 70 volumes, they were averaged and only the mean volume was segmented;
this reduced the data size and improved the SNR. In our case, the
structure of interest was the inner segment/outer segment junction,
which is a relatively smooth surface. Therefore, it is possible to
coarse grain the volume of size $640\times368\times256$ to further
reduce data size. In our case $10\times10\times1$ voxels were concentrated
in one single voxel; the value of this resulting voxel was computed
as the average intensity of this area. Afterwards, for each A-scan
the four highest local maxima were extracted, which are characterized
by their $x$, $y$, and $z$ coordinate. Given these data points,
we created a transition probability matrix between any two points
that are separated by $\Delta x$, $\Delta y$, and $\Delta z$ (measured
in units of pixels). It is given by 

\[
\textbf{p}=\begin{cases}
G_{\sigma_{r},\sigma_{\theta}}\left(r,\theta\right)\text{,} & \Delta x\le1\text{ and }\Delta y\le1\text{ and }\Delta z\le1\\
0 & \text{otherwise}
\end{cases}
\]
where $r=\sqrt{\Delta x^{2}+\Delta y^{2}+\Delta z^{2}}$, $\theta=\cos^{-1}\left(\Delta z/r\right)$,
and $G_{\sigma_{r},\sigma_{\theta}}$ is a Gaussian function in two
dimensions with covariance matrix $\mathrm{diag}\left(\sigma_{r}^{2},\sigma_{\theta}^{2}\right)$.
With this matrix, transitions are only allowed for neighboring data
points and lateral transitions are more likely than axial transitions.
In a next step, a diffusion map of these data points was calculated
\cite{coifman}. The diffusion map maps the original data points to
a new coordinate system depending on the transition probability matrix;
in the new coordinate system points are closer together, if they have
a high transition probability after a specified number of transitions
$N$. We used $N=10,000$ to reach a stationary state. Finally, a
$k$-means clustering of the data points based on their distance in
the diffusion coordinate system is done giving clusters, one of which
corresponds to the IS/OS layer. The cluster of the IS/OS were used
for a surface fit. Afterwards, each A-scan in all volumes is axially
shifted by the Fourier shift theorem to align the IS/OS-surface to
a constant depth.

\paragraph{Phase evaluation}

The phases in recorded volumes do not carry information about absolute
position and can only measure changes when compared to phases in other
layers and at other times. To cancel this arbitrary phase offset in
each pixel the reconstructed volumes were first referenced to a volume
before the start of the optical stimulus. For the evaluation of the
phase differences the complex OCT signal was averaged over several
layers. Changes in the OPL of the photoreceptor outer segments were
calculated from the phase difference of the inner segment/outer segment
junction (averaged over 2 layers) and the outer segments tips (averaged
over 2 layers) located 4 pixel deeper for cones and 6 pixel deeper
for rods. For the evaluation of the ganglion cells the phase difference
of the GCL (averaged over 6 pixels) and the IPL (averaged over 5 pixel)
was calculated. The layer thickness of the GCL and IPL varies with
the lateral position in the retina. Therefore the distance between
those two layers is not fixed and needed to be chosen manually for
each position individually (Fig.\ \ref{time_course}). Since retinal
vessels are in the depth of the GCL, the phase difference are dominated
by motion artifacts due to pulsation of the vessels. To minimize those
artifacts each volume was referenced to one of the five volumes acquired
before stimulation, which provides the smallest phase noise. This
is the case if the reference volume is in a similar phase of the retinal
heart-beat induced pulsation. The phase error for each possible reference
volume was therefore determined by the standard deviation of the phase
difference histogram. The image quality was improved by applying a
lateral Gaussian filter to the complex data. Averaging over the whole
stimulation time (5\textsuperscript{th} - 70\textsuperscript{th}
volume) further improved the image quality of the response of the
GCL (Fig.\ \ref{translation}b). The images of the response of the
photoreceptor OS were calculated only from the 20\textsuperscript{th}
volume (after 1875 ms of stimulation). The time-courses of the response
of rods and cones were calculated from areas, which were selected
manually (Fig.\ \ref{translation}a). Averaging over this area improved
signal quality. For phase changes larger than $\pi$ the phase was
unwrapped. OPL changes were calculated by

\[
\Delta\ell=\frac{\Delta\Phi}{4\pi}\lambda_{0.}
\]

In the GCL the time-course of the IOS was corrupted by inhomogeneously
varying background changes in the phase, which were not connected
to the optical stimulation. This background changes were removed by
manually masking the area were the IOS arose and the vessels dominated
the phase. The phase in the remaining background area was averaged,
unwrapped and subtracted from the time course received from the IOS,
before it was rescaled to length information.

To create the lateral translation map between photoreceptors and ganglion
cells (Fig.~\ref{translation}c), the stimulation response in each
image was outlined manually. Corresponding corners of the cross in
each outline were then connected giving the results shown in Fig.~\ref{translation}c.

\paragraph*{Data Availability }

The data that support the findings of this study are available from
the corresponding author upon reasonable request.

\bibliographystyle{naturemag}
\bibliography{Pfaeffle_FunctionalImaging}

\end{document}